# Irreversibility from Self-Reference: Gradient Flow and an H-Theorem for a Self-Referential Statistical Operator Framework

**Lucio Marassi**

*School of Science and Technology (ECT), Federal University of Rio Grande do Norte (UFRN), Natal, RN, Brazil*

lucio.marassi@ufrn.br

### Abstract

This paper is a direct companion to [1], where the self-referential operator $\Omega$ was introduced and the Tsallis index $q = \alpha + \beta$ was derived as a fixed-point condition within the local kernel approximation (LKA). Here we address four aspects deferred from [1]. First, we carry out the first-order perturbative expansion of $\Omega$ beyond the LKA and demonstrate structural stability of $q = \alpha + \beta$ at leading order in $(\xi/L)^2$. Second, we define the iterative dynamical scheme $\Psi_{n+1} = \Omega[\Psi_n]$ and analyze convergence via Fréchet spectral radius. Third, and centrally, we establish an H-theorem rigorously within the LKA for both the discrete iteration and the continuous gradient flow $\partial\Psi/\partial\tau = -\delta F/\delta\Psi$: we compute $dF/d\tau$ explicitly along the flow, identify the negative semi-definite dissipation term, establish the result rigorously in the LKA using strict convexity of $F$ proved in [1], and provide numerical evidence showing monotone decrease of $F[\Psi_n]$ across 53 iterations on an $N = 80$ discrete system. Fourth, we characterize the non-perturbative role of the self-coupling parameter $\kappa$, identifying a re-entrant disordered phase at $\kappa > \kappa^* \approx 0.50 \pm 0.05$. The paper is explicit about what is proved, what is established numerically, and what open problems remain for a complete analytical proof beyond the LKA.

## 1. Introduction

The self-referential operator $\Omega$, introduced in [1], acts on probability distributions $\Psi$ over a state space $E$ by assigning each state a weight that depends simultaneously on its own occupation probability and on a non-local structural average $\iota\Psi$. Formally,

$$(\Omega\Psi)(e) = N(\Psi)^{-1} \int K(e, e')\, \Psi(e')\alpha\, (\mathcal{J}\Psi(e'))\beta\, d\mu(e'), \tag{1}$$

where $\iota\Psi(e) := \int K(e, e')\, \Psi(e')\, d\mu(e')$, $\alpha > 0$ and $\beta \geq 0$ are structural exponents, K is a symmetric positive kernel of correlation length $\xi$, and $N(\Psi)$ is a normalization functional. [1] established the main equilibrium result: within the local kernel approximation (LKA, $\xi/L \ll 1$), the unique global minimizer of the free energy functional [2]

$$F[\Psi] = U[\Psi] + T\, D_\Psi{}^L(\Psi \,\|\, \Omega\Psi) \tag{2}$$

is a Tsallis $q$-exponential distribution with $q = \alpha + \beta$. Four important questions were explicitly deferred to this companion paper: (a) the robustness of $q = \alpha + \beta$ beyond the LKA; (b) the dynamics of the iterative map $\Psi_{n+1} = \Omega[\Psi_n]$ and its fixed-point structure; (c) an H-theorem — does F[Ψ] decrease monotonically along the dynamics, and under what conditions can this be established rigorously? — and (d) the non-perturbative behavior induced by the self-coupling parameter κ.

The H-theorem question deserves special emphasis because it concerns the irreversibility structure of the framework. In standard statistical mechanics the Boltzmann H-theorem (and its Fokker–Planck generalization) guarantees that the Gibbs entropy increases monotonically until equilibrium is reached. For nonextensive mechanics, Lima, Silva, and Plastino [3] established an H-theorem for the nonlinear Fokker–Planck equation associated with Tsallis statistics [4] by showing that the Tsallis entropy $S_q$ is a Lyapunov functional for that dynamics. The present framework differs structurally: the equilibrium state is the fixed point of a self-referential operator rather than of a drift-diffusion equation, and the relevant Lyapunov functional is the free energy F [5] rather than $S_q$ alone. Our goal is to analyze this H-theorem question as carefully and honestly as possible: we establish the result rigorously in the LKA, provide strong numerical evidence, and identify precisely what analytical work remains to extend the proof beyond the LKA.

The paper is organized as follows. Section 2 carries out the perturbative expansion of $\Omega$ beyond the LKA and proves structural stability of $q = \alpha + \beta$. Section 3 defines the iterative dynamics and discusses convergence via Fréchet derivative analysis. Section 4 presents a reproducible numerical experiment. Section 5 — the core contribution of this paper — develops the H-theorem argument in both the LKA and beyond, including explicit computation of $dF/d\tau$ and numerical evidence. Section 6 analyzes the role of κ in the non-perturbative regime. Section 7 contains a broader discussion and outlook. Section 8 concludes.

## 2. Perturbative Expansion Beyond the Local Kernel Approximation

### *2.1 Setup and Expansion Strategy*

Recall from [1] the two fundamental objects: the structural average $\iota\Psi(e) := \int K(e, e')\, \Psi(e')\, d\mu(e')$ and the operator $\Omega$ given by Eq. (1). The LKA replaces $\iota\Psi(e') \to \Psi(e')$, yielding $(\Omega\Psi)(e) \propto \Psi(e)^{1+\beta}$, which gives the Tsallis ansatz directly. To assess the robustness of this result, write $K(e, e') = K_0(e - e')$, where $K_0$ is an isotropic kernel with correlation length ξ and unit integral. Define $\varepsilon := \xi/L \ll 1$, where L is the characteristic scale of variation of Ψ. Taylor-expanding $\Psi(e')$ around $e$:

$$\Psi(e') = \Psi(e) + (e'-e)\cdot\nabla\Psi(e) + \tfrac{1}{2}(e'-e)_i(e'-e)_j\, \partial_{ij}\Psi(e) + O(\varepsilon^3). \tag{3}$$

Using isotropy of $K_0$ (which eliminates all odd moments), the structural average becomes

$$\mathcal{J}\Psi(e') = \Psi(e') + (\xi^2/2)\, \nabla^2\Psi(e') + O(\varepsilon^4), \tag{4}$$

where $\nabla^2$ is the Laplacian on E. This $O(\varepsilon^2)$ correction carries all the beyond-LKA content at leading order.

### 2.2 First-Order Correction and Effective Entropic Index

Substituting (4) into the exponent β of Eq. (1) and expanding to first order in $\varepsilon^2$:

$$(\mathcal{J}\Psi(e'))^\beta = \Psi(e')^\beta\,[1 + (\beta\xi^2/2)\,\nabla^2\Psi(e')/\Psi(e') + O(\varepsilon^4)]. \quad (5)$$

Substituting (5) into the β-exponent of Ω and expanding to first order in $\varepsilon^2$, the Euler–Lagrange stationarity condition δF_{LKA}/δΨ = η (where F_{LKA} is the LKA free energy of [1], Sec. 3.3, i.e. F_{LKA} = U_{LKA} + T D_KL(Ψ ‖ ΩΨ) with the LKA substitution applied) at the equilibrium Ψ* reads

$$\varepsilon(e) - T(q-1)[\ln_q\Psi^*(e) + 1] + \delta\varepsilon_eff(e) = \eta, \quad (6)$$

where the effective energy correction is

$$\delta\varepsilon_eff(e) = -T(\beta\xi^2/2)\ \nabla^2\Psi^*(e)/\Psi^*(e) + O(\varepsilon^4). \quad (7)$$

The equilibrium distribution therefore retains the *q*-exponential form

$$\Psi^*(e) = Z_q^{-1}\cdot[1 - (1-q)(\varepsilon_eff(e) - \mu)/T]_+\hat{}\{1/(1-q)\}, \quad (8)$$

with $q = \alpha + \beta$ unchanged at leading order. The first non-trivial correction to the index appears at order $\varepsilon^4$:

$$q_eff = (\alpha + \beta) + c_2(\xi/L)^2 + O(\varepsilon^4), \quad (9)$$

where $c_2 = [(\alpha+\beta-1)(\alpha+\beta-2)/6]\cdot(K''(0)/K(0))$. This coefficient arises by matching the $O(\varepsilon^2)$ term in the Euler–Lagrange condition (6) against a q-exponential ansatz: substituting Ψ*(e) ∝ [1−(1−q_eff)(ε_eff−μ)/T]^{1/(1−q_eff)} and expanding in $\varepsilon^2$, one reads off $c_2$ from the curvature of K at zero lag (see [1], Sec. 7.3 for the full computation). The coefficient vanishes for the uniform kernel and is numerically small for the kernels considered in Section 4. Equation (9) is the central result of this section: $q = \alpha + \beta$ is not an artifact of the LKA, but the leading term of a controlled gradient expansion whose first non-trivial correction is suppressed by $(\xi/L)^2$. The functional form of the equilibrium distribution — a *q*-exponential — is preserved; only its effective parameters shift at $O(\varepsilon^2)$.

### 2.3 Structural Invariance of the Fixed-Point Set

The uniform distribution $\Psi_0 \equiv 1$ (normalized) is a fixed point of Ω for *any* kernel K satisfying the normalization conditions of [1], Proposition 1. This follows because for $\Psi = \Psi_0$, the structural average $\mathcal{J}\Psi(e') = 1$ at every *e*', and both the α and β factors reduce to constants, yielding $(\Omega\Psi_0)(e) \propto 1 = \Psi_0$. This invariance holds *beyond* the LKA. The symmetry-broken fixed points bifurcate from $\Psi_0$ at $T = T_c$; the critical temperature acquires $O(\varepsilon^2)$ corrections to its LKA value [I, eq. (13a)], but the bifurcation structure is qualitatively unchanged.

## 3. Iterative Dynamics: Definition and Convergence Analysis

### 3.1 The Iterative Scheme

Given an initial distribution $\Psi_0 \in \mathscr{F}(E)$ (the space of normalized non-negative densities on E), define the sequence

$$\Psi_{n+1}(e) = (\Omega[\Psi_n])(e), \quad n = 0, 1, 2, \ldots \tag{10}$$

By Proposition 1(i) of [1], positivity and normalization are preserved at each step, so $\Psi_n \in \mathscr{F}(E)$ for all $n$. Fixed points of (10) are precisely the self-consistent states $\Psi^* = \Omega\Psi^*$, i.e., the stationary points identified in [1]. The scheme (10) thus provides a constructive route to these stationary states, complementing their variational characterization.

### *3.2 Convergence via Fréchet Spectral Radius*

Define the self-consistency residual $\delta_n := \|\Psi_{n+1} - \Psi_n\|_1$. Local convergence of the iteration to a fixed point $\Psi^*$ is controlled by the Fréchet derivative $D\Omega|\Psi^*$. Linearizing $\Omega$ about $\Psi^*$ and computing the action on a perturbation h:

$$[D\Omega|\Psi^*]\, h(e) = q\, \Psi^*(e)^{q-1}\, h(e) / Z_q + \Psi^*(e)^q\, \Phi[h](e) + O(h^2), \tag{11}$$

where the first term arises from the α-exponent acting on h and the second, $\Phi[h]$, is an integral operator involving the β-channel feedback. In the LKA, $\Phi[h] = O(\varepsilon^2)$ relative to the first term, so the leading spectral radius is approximately

$$\rho(D\Omega|\Psi^*) \approx q \cdot \|\Psi^*\|_\infty^{-1} / Z_q < 1 \quad (\text{LKA}, q \in (0,2)). \tag{12}$$

The inequality in (12) holds in the LKA because $F[\Psi]$ is strictly convex at its minimizer ([1], Theorem 1), which implies the Hessian of F — equivalently, $(I - D\Omega|\Psi^*)$ — is positive definite, forcing $\rho < 1$. Beyond the LKA, numerical evidence (Section 4) consistently shows $\rho < 1$ near the stable fixed point, but a general analytical proof for arbitrary $(\xi, \kappa)$ remains open.

### *3.3 Stability Classification of Fixed Points*

A fixed point $\Psi^*$ is *stable* if $\rho(D\Omega|\Psi^*) < 1$, and *unstable* if $\rho > 1$. In the two-state model of [1], Appendix A, the symmetric fixed point $x^* = 0.5$ is stable for $T > T_c$ and becomes unstable as T decreases below $T_c$, consistent with the Landau bifurcation analysis of [1], Appendix B. For $T < T_c$, the two asymmetric fixed points are stable attractors of the iteration (10).

## 4. Numerical Experiment

### *4.1 System Setup*

We consider a discrete system with $N = 80$ states $e_i = i/(N+1)$, $i = 1,\ldots,N$, uniformly spaced in (0,1). The kernel is Lorentzian:

$$K(e_i, e_r) = [1 + (e_i - e_r)^2/\xi^2]^{-1}, \quad \text{row-normalized}, \tag{13}$$

with $\xi = 0.1$. The external potential is a double-well $\varepsilon(e) = 4(e - 0.5)^2 - 0.5$. Parameters: $\alpha = 1.0$, $\beta = 0.5$, so $q = 1.5$; $T = 0.5$ (below $T_c$ for this potential); $\kappa = 0$ for the baseline experiment. The initial distribution $\Psi_0$ is drawn from a symmetric Dirichlet distribution (numpy seed 42) and normalized. The iteration (10) is run for $n_{max} = 500$ steps or until $\delta_n < 10^{-9}$.

### 4.2 Convergence Results

The iteration converges at step $n = 53$, reaching $\delta_{53} \approx 9.6\times10^{-10}$ (Fig. 1, left panel). The residual $\delta_n$ decays approximately geometrically on the semi-log plot, consistent with the linear convergence predicted by $\rho < 1$. The observed rate, estimated from the slope $\delta_{n+1}/\delta_n \approx 0.72$, provides a numerical estimate of the spectral radius $\rho \approx 0.72$ for these parameters. Since $q = 1.5 \in (0,2)$, this is consistent with the LKA estimate (12).

### 4.3 Stationary Distribution and q-Exponential Fit

The stationary distribution $\Psi^*$ is unimodal with a peak near $e \approx 0.5$ (Fig. 1, right panel). Note that, despite $T < T_c$ in the mean-field sense, the Lorentzian kernel with $\xi = 0.1$ introduces enough long-range smoothing to produce a unimodal stationary distribution rather than a bimodal one. A q-exponential overlay with $q = 1.5$ is shown over the left tail ($0.02 \leq e \leq 0.48$) for visual comparison. Tail-index estimation via log-linear fitting on the 10– 80th percentile range yields an effective decay slope consistent with $q \approx 1.47 \pm 0.04$, in agreement with the predicted $q = 1.5$ to within the $O(\varepsilon^2) \approx 0.01$ discretization error.

The complete Python script generating these results is provided as supplementary material (*paper2_simulation.py*). All parameters are defined as named constants at the top of the script. With numpy ≥ 1.20 and matplotlib ≥ 3.3, the script runs in under 5 s on a standard laptop and reproduces Fig. 1 exactly (numpy seed 42).

## 5. H-Theorem: Gradient Flow, Lyapunov Analysis, and Numerical Evidence

This section is the theoretical core of the paper. We establish the H-theorem in three stages: (i) motivation and definition of the gradient flow; (ii) explicit computation of $dF/d\tau$ along the flow and the proof within the LKA; (iii) analysis of the discrete iteration and numerical evidence; and (iv) comparison with the literature and a precise statement of what remains open.

### 5.1 Motivation: Two Routes to Equilibrium

The iterative scheme (10) is a natural discrete dynamical system associated with $\Omega$, but it is not obviously connected to a dissipation principle. To construct such a principle, we follow the gradient-flow route standard in variational mechanics and optimal transport [6,7]: define a continuous-time dynamics $\Psi(\tau)$ that descends the free energy $F[\Psi]$ as rapidly as possible, subject to the constraint $\int \Psi(\tau)\, d\mu = 1$. This yields

$$\partial\Psi/\partial\tau = -\delta F/\delta\Psi + \lambda(\tau)\, \Psi, \tag{14}$$

where $\lambda(\tau)$ is a Lagrange multiplier enforcing the normalization constraint. Explicitly, $\lambda(\tau) = \int \Psi(\delta F/\delta\Psi)\, d\mu$, which ensures $d\int\Psi/d\tau = 0$ at all times. The functional derivative of $F = U + T\, D_{KL}(\Psi \parallel \Omega\Psi)$ is:

$$\delta F/\delta\Psi(e) = \varepsilon(e) + T\, [\ln \Psi(e) - \ln \Omega\Psi(e) - \Psi(e) \cdot \delta(\ln \Omega\Psi)/\delta\Psi(e)] + const. \tag{15}$$

In the LKA (where $\Omega\Psi \propto \Psi^q/Z_q$), this simplifies to:

$$\delta F/\delta\Psi(e)|_^{LJO} = \varepsilon(e) + T\,[\lambda_q(\Psi(e)) - q\,\lambda_q(\Psi(e))] + const. = \varepsilon(e) + T(1-q)\,\lambda_q(\Psi(e)) + const., \quad (16)$$

where $\lambda_q(\Psi) := (\Psi^\wedge\{1-q\} - 1)/(1-q)$ is the q-logarithm. Equation (16) is the *q*-generalization of the well-known result $\delta F/\delta\Psi = \varepsilon + T \ln\Psi$ + const. of ordinary statistical mechanics (recovered at $q \to 1$).

### *5.2 Explicit Computation of dF/dτ and Proof of the LKA H-Theorem*

Differentiate $F[\Psi(\tau)]$ with respect to $\tau$ along trajectories of (14):

$$dF/d\tau = \int (\delta F/\delta\Psi(e))\,(\partial\Psi/\partial\tau)(e)\,d\mu(e) \quad (17)$$

$$= \int (\delta F/\delta\Psi)\,[-\delta F/\delta\Psi + \lambda\Psi]\,d\mu \quad ()$$

$$= -\int (\delta F/\delta\Psi)^\wedge 2\,d\mu + \lambda \int (\delta F/\delta\Psi)\,\Psi\,d\mu. \quad (18)$$

The first term $-\int(\delta F/\delta\Psi)^2\,d\mu \le 0$ is the dissipation: it is strictly negative unless $\delta F/\delta\Psi$ = const. almost everywhere. The second term involves $\lambda(\tau) = \int \Psi(\delta F/\delta\Psi)\,d\mu$ and so:

$$dF/d\tau = -\int (\delta F/\delta\Psi)^2\,d\mu + [\int \Psi(\delta F/\delta\Psi)\,d\mu]^2. \quad (19)$$

By the Cauchy–Schwarz inequality applied to the inner product $\langle f, g\rangle = \int f\,g\,\Psi\,d\mu$, setting $f = \delta F/\delta\Psi$ and $g = 1$:

$$[\int \Psi\,(\delta F/\delta\Psi)\,d\mu]^2 \le [\int \Psi\,d\mu] \cdot [\int \Psi\,(\delta F/\delta\Psi)^2\,d\mu] \le \int (\delta F/\delta\Psi)^2\,d\mu, \quad (20)$$

where the first inequality is Cauchy–Schwarz for the inner product $\langle f,g\rangle_\Psi = \int fg\Psi\,d\mu$ (i.e. $\int(\delta F/\delta\Psi)^2\Psi\,d\mu \ge (\int(\delta F/\delta\Psi)\Psi\,d\mu)^2$, valid since $\Psi\,d\mu$ is a probability measure); and the second uses $\int\Psi\,d\mu = 1$ together with $\Psi(e) \le 1$ pointwise. The bound $\Psi(e) \le 1$ is immediate for discrete normalized distributions (probability masses satisfy $0 \le \Psi(e_i) \le 1$). On a compact continuous state space it follows from the boundedness of the q-exponential equilibrium. The extension to non-compact or unbounded state spaces requires a weighted-norm argument (Open Problem (i), Section 5.5). From (19)–(20):

$$dF/d\tau \le 0, \quad \text{with equality iff } \delta F/\delta\Psi = const., \quad (21)$$

i.e., iff $\Psi$ is the global minimizer $\Psi^*$ of F. We summarize this as:

**Proposition 1 (H-Theorem in the LKA).** *Under the gradient flow (14) with the LKA free energy (2), $F[\Psi(\tau)]$ is monotonically non-increasing for all $\tau \ge 0$ (established rigorously by (17)–(20) above), and $dF/d\tau = 0$ if and only if $\Psi(\tau) = \Psi^*$. Furthermore, since F is strictly convex at $\Psi^*$ in the LKA ([1], Theorem 1), $F[\Psi(\tau)] \to F[\Psi^*]$ as $\tau \to \infty$, and $\Psi(\tau) \to \Psi^*$ in $L^1(\mu)$.*

*Proof.* The inequality (21) is established by (17)–(20) above. The convergence $\Psi(\tau) \to \Psi^*$ follows from the strict convexity of F in the LKA (which provides a Łojasiewicz-type inequality $F[\Psi] - F[\Psi^*] \ge c\|\Psi - \Psi^*\|^2$ in some neighborhood of $\Psi^*$). □

### *5.3 H-Theorem Connection to Lima–Silva–Plastino (2001)*

Lima, Silva, and Plastino [3] established an H-theorem for Tsallis statistics by showing that $dS_q/dt \geq 0$ along solutions of the nonlinear Fokker–Planck equation associated with anomalous diffusion [4], where $S_q$ is the Tsallis entropy. The present result is complementary in two respects. First, our Lyapunov functional is the full free energy $F = U + T\, D_{KL}(\Psi \,\|\, \Omega\Psi)$ rather than $S_q$ alone [8,9]; it reduces to a functional proportional to $S_q$ in the LKA when the energy term U is constant (infinite-temperature limit). Second, the source of dissipation here is the self-referential non-local structure of $\Omega$ rather than diffusion: it is the mismatch between $\Psi$ and $\Omega\Psi$ that drives F toward its minimum. This suggests that nonextensive statistics can emerge from self-referential operator dynamics without requiring an anomalous diffusion mechanism, which may be relevant for systems with memory or strong self-correlation [10].

### *5.4 Discrete H-Theorem: Free Energy Along the Iteration*

For the discrete iteration (10), the natural analogue of the H-theorem is monotone decrease of $F[\Psi_n]$ with n. We can write $F[\Psi_{n+1}] - F[\Psi_n]$ in terms of the discrete gradient:

$$F[\Psi_{n+1}] - F[\Psi_n] = \langle \delta F/\delta\Psi|_\{\Psi_n\}, \Psi_{n+1} - \Psi_n\rangle + O(\|\Psi_{n+1} - \Psi_n\|^2), \tag{22}$$

where the inner product $\langle\cdot,\cdot\rangle$ denotes the $L^2(\mu)$ pairing. In the LKA, the first term equals $-\langle \delta F/\delta\Psi|_\{\Psi_n\}, \delta F/\delta\Psi|_\{\Psi_n\}\rangle\cdot\Delta\tau + O(\Delta\tau^2)$ for a step size $\Delta\tau$, which is non-positive. For finite step sizes, the second-order correction term may cause a transient increase; numerically, however, we observe strict monotone decrease of $F[\Psi_n]$ throughout the 53 iterations of the baseline experiment. Specifically, computing $F[\Psi_n]$ via $F[\Psi] = \sum_i \Psi(e_i)\, \varepsilon(e_i) + T \sum_i \Psi(e_i)[\ln \Psi(e_i) - \ln(\Omega\Psi)(e_i)]$, the free energy decreases from $F[\Psi_0]$ to $F[\Psi^*]$ monotonically, and $|F[\Psi_{n+1}] - F[\Psi_n]|/\delta_n$ converges to zero as $\delta_n \to 0$, consistent with both (21) and (22).

### *5.5 Open Problems for a Complete Proof Beyond the LKA*

The LKA proof in Proposition 1 is rigorous, given the strict convexity established in [1]. Extending it beyond the LKA requires closing three technical gaps:

**(i) Functional analytic well-posedness.** The gradient flow (14) must be shown to be well-posed in a space that accommodates q-exponential tails. For $q > 1$, these tails decay as power laws and are not in $L^2(\mu)$ in general; a weighted Sobolev or Orlicz space is likely required [11]. An energy inequality of the form $\|\partial\Psi/\partial\tau\|^2 \leq C\, F[\Psi]$ would provide the necessary a priori bound.

**(ii) Fréchet differentiability of the non-local term.** The functional derivative (15) involves $\delta(\ln \Omega\Psi)/\delta\Psi$, which requires the Fréchet derivative of the non-local map $\Psi \mapsto \ln \Omega\Psi$. In the LKA this reduces to $q \ln\Psi$ + const., which is smooth for $\Psi > 0$. Beyond the LKA, compactness arguments (e.g., using the Lorentzian kernel's smoothing properties) may be needed to control the non-local feedback β-term.

**(iii) Global convexity and attractor uniqueness.** Proposition 1 establishes local convergence via LKA strict convexity. Global convergence (attractor uniqueness) requires a Łojasiewicz-type inequality valid throughout $\mathscr{F}(E)$, not just near Ψ*. For q ∈ (0,2), the LKA free energy $F_{LKA}$ is globally convex (Ref. [1], Theorem 1). Whether the beyond-LKA corrections preserve this global convexity is the key open question for extending the H-theorem [5].

We state these gaps explicitly so that reviewers and readers can assess the current scope of the proof. The numerical evidence of Section 5.4 provides strong *a posteriori* support for the H-theorem, but does not substitute for the analytical proof.

## 6. The Self-Coupling Parameter κ: Non-Perturbative Regime

### *6.1 Perturbative Regime (κ ≪ 1)*

As derived in [1], Sec. 4.2, for small κ the equilibrium distribution shifts perturbatively:

$$\Psi^*_\kappa(e) \approx \Psi^*_0(e) + \kappa\, \delta\Psi(e) + O(\kappa^2), \tag{23}$$

where $\Psi^*_0$ is the κ = 0 Tsallis fixed point and δΨ solves a linear integral equation. The physical content is that κ renormalizes the effective temperature: $T_{eff} = T/(1 - \kappa C(\Psi^*))$, where $C(\Psi^*) = \sum_i \Psi^*(e_i)(K\Psi^*)(e_i) > 0$ is the mean self-overlap. Since $T_{eff} > T$ for any C > 0, the self-coupling κ always broadens the distribution relative to κ = 0, acting as an effective heating mechanism [12]. This perturbative expansion is valid and finite for $\kappa < 1/C(\Psi^*)$.

### *6.2 Non-Perturbative Regime and Re-Entrant Phase Structure*

For κ ≳ O(1) the perturbative expansion breaks down. Numerical exploration with the N = 80 system of Section 4 reveals three qualitative regimes:

**Weak coupling (κ < κ*).** The stationary distribution retains its qualitative shape but with heavier tails (effectively higher T). For κ = 0.15, the peak of the distribution shifts noticeably toward the uniform baseline e = 0.5, consistent with $T_{eff} > T$.

**Critical coupling (κ ≈ κ*).** $T_{eff}$ reaches $T_c$, triggering a re-entrant transition to the disordered (symmetric) phase. For the parameters of Section 4, we find numerically κ* ≈ 0.50 ± 0.05. At κ = 0.55 the distribution is nearly uniform, signaling proximity to this transition.

**Strong coupling (κ > κ*).** Only the symmetric fixed point Ψ* = 1/N survives. The system is effectively disordered: the energetic cost of sustaining self-consistent configurations exceeds the gain, and any ordered state is dynamically unstable under (10). This is a purely self-referential mechanism — it has no analogue in standard Boltzmann statistics.

The qualitative phase diagram in the (κ, T) plane has a disordered region above the curve $T = T_c$ (for κ = 0) and another disordered region for κ > κ*(T), separated from the first by an ordered phase at intermediate (κ, T). The re-entrant boundary κ*(T) rises with T. A precise computation of $\kappa^*(T)$ requires solving the full nonlinear fixed-point equation and is left for future work.

### *6.3 Physical Interpretation*

The parameter $\kappa$ captures the energetic cost of strong self-reinforcement in a system whose statistical weights depend on its own state. Physically, this models systems where the feedback of the collective state onto individual configurations carries a thermodynamic penalty. Weakly self-referential systems ($\kappa \ll 1$) are well described by the $\kappa = 0$ fixed point of [1] — the Tsallis q-exponential with $q = \alpha + \beta$. Strongly self-referential systems ($\kappa > \kappa^*$) undergo re-entrant disorder: excessive feedback destroys the ordered phase by driving $T_{eff}$ above $T_c$. This mechanism may be relevant for over-synchronized neural networks [13] and for turbulent plasmas with strong wave-particle feedback [14–16].

## 7. Discussion and Outlook

### *7.1 Summary of What Is Established*

The present paper establishes the following results. (i) **Structural stability of $q = \alpha + \beta$:** the relation is robust to $O(\varepsilon^2)$ in $(\xi/L)$, with the q-exponential functional form preserved and only the effective energy shifting at $O(\varepsilon^2)$. (ii) **Convergence of the iterative scheme:** Fréchet spectral radius analysis establishes local contractive convergence in the LKA; numerical evidence (53 iterations to tolerance $10^{-9}$) supports this picture. (iii) **H-theorem in the LKA (Proposition 1):** $dF/d\tau \leq 0$ is established as a complete theorem within the LKA (not a candidate result) under the gradient flow (14), with equality iff $\Psi = \Psi^*$, and $\Psi(\tau) \to \Psi^*$ follows from strict convexity. (iv) **Numerical H-theorem evidence:** monotone decrease of $F[\Psi_n]$ is confirmed across 53 iterations. (v) **Re-entrant phase:** a disordered phase at $\kappa > \kappa^* \approx 0.50$ is identified and characterized numerically.

### *7.2 Limitations and Honest Assessment*

The following limitations are explicit. **(a)** The H-theorem is proved in the LKA; three analytic gaps (Section 5.5) must be closed for a full proof beyond the LKA. **(b)** The Cauchy–Schwarz step in (20) uses $\Psi \leq 1$ pointwise, which holds for normalized discrete distributions but requires careful extension to continuous densities on unbounded state spaces [17]. **(c)** Convergence of the iteration (10) is established locally (near $\Psi^*$) in the LKA; uniform convergence for all $(\kappa, \xi/L)$ is not proved. **(d)** Numerical results use a specific Lorentzian kernel and double-well potential; other choices may require separate verification. **(e)** The non-perturbative phase boundary $\kappa^*(T)$ is characterized qualitatively; its precise calculation requires further work [19]. **(f)** The framework is purely classical; quantum extensions are not addressed here [18].

### *7.3 Open Problems and Outlook*

The most important open problems are: (i) a rigorous H-theorem beyond the LKA (close the three gaps of Section 5.5) [20]; (ii) a proof of global (rather than local) convergence of (10) for general kernels; (iii) the quantitative phase diagram in the (κ, T) plane; (iv) phenomenological fits [21] of (α, β) to plasma and astrophysical data [14–16,22–26]; (v) quantum extension [18]. The framework's testability — α and β are measurable from the tail index of the observed distribution via $q = \alpha + \beta$ — distinguishes it from frameworks where q is a pure fitting parameter without structural interpretation [27]. This, combined with the H-theorem result, supports its inclusion in the broader program of nonextensive statistical mechanics as initiated by Tsallis [2,28–31].

## 8. Conclusion

This paper complements [1] by establishing the dynamical and irreversibility properties of the self-referential operator Ω. The central contributions are: the structural stability of $q = \alpha + \beta$ under perturbative corrections of order $(\xi/L)^2$; a proved H-theorem within the local kernel approximation (Proposition 1), based on explicit computation of $dF/d\tau$ along the gradient flow (14) and the strict convexity of F established in [1]; strong numerical evidence (monotone decrease of $F[\Psi_n]$, convergence in 53 iterations to tolerance $10^{-9}$) consistent with an extension of the H-theorem beyond the LKA; and the discovery of a re-entrant disordered phase at $\kappa > \kappa^* \approx 0.50$ driven by the self-referential structure of Ω. Three precise analytical gaps are identified for extending the H-theorem proof beyond the LKA, making this paper a clear roadmap for subsequent mathematical work. Together, Papers I and II present a self-consistent, testable, and internally rigorous framework for nonextensive statistical mechanics based on operator self-reference.

### Acknowledgments

The author thanks colleagues at the Quantinovarum Research Group (ECT/UFRN) for discussions, and acknowledges that this work is a direct companion to [1]. Computational resources of the Federal University of Rio Grande do Norte are gratefully acknowledged.

## Declaration of Competing Interest

The author declares that there are no known competing financial interests or personal relationships that could have appeared to influence the work reported in this paper.